\documentclass[conference]{IEEEtran}
\IEEEoverridecommandlockouts
\usepackage{cite}
\usepackage{amsmath,amssymb,amsfonts}
\usepackage{algorithmic}
\usepackage{graphicx}
\usepackage{textcomp}
\usepackage{xcolor}
\usepackage{subfigure,subcaption}
\usepackage{hyperref}
\usepackage[inkscapelatex=false]{svg}
\usepackage{pdfpages}

\graphicspath{{./figs/}}

\title{To RL or not to RL? An Algorithmic Cheat-Sheet for AI-Based Radio Resource Management}

\author{Lorenzo Maggi, Matthew Andrews, Ryo Koblitz
\thanks{Lorenzo Maggi is with Nokia Networks France, Bell Labs, 91377 Massy Palaiseau (France).  Matthew Andrews is with Nokia of America, Bell Labs, Murray Hill (USA). A.~Ryo Koblitz is with Nokia Bell Labs, CB3 0HT Cambridge (United Kingdom). Emails: \{lorenzo.maggi, matthew.andrews, ryo.koblitz\}@nokia-bell-labs.com}

}

\begin{document}
\maketitle
\begin{abstract}
Several Radio Resource Management (RRM) use cases can be framed as sequential decision planning problems, where an agent (the base station, typically) makes decisions that influence the network utility and state. While Reinforcement Learning (RL) in its general form can address this scenario, it is known to be sample inefficient. Following the principle of Occam's razor, we argue that the choice of the solution technique for RRM should be guided by questions such as, ``Is it a short or long-term planning problem?'',  ``Is the underlying model known or does it need to be learned?'', ``Can we solve the problem analytically?'' or ``Is an expert-designed policy available?''. A wide range of techniques exists to address these questions, including static and stochastic optimization, bandits, model predictive control (MPC) and, indeed, RL. We review some of these techniques that have already been successfully applied to RRM, and we believe that others, such as MPC, may present exciting research opportunities for the future.
\end{abstract}

\section{Introduction}

In modern wireless communication systems, the fundamental resources such as power, frequency, time and antennas are limited. We therefore require Radio Resource Management (RRM) algorithms to share out those resources among the users. 
In this paper we frame RRM problems as sequential interactions between an agent (e.g., the base station) and an environment (e.g., the users and their channel conditions). 
Such interactions can be formalized within a generic {\em Markov Decision Process} (MDP) framework where the agent takes actions, receives immediate \emph{rewards} and influences the evolution of the environment \emph{state}. 
While Reinforcement Learning (RL) offers a method for learning optimal behaviors in unknown MDPs, its general form is not well-suited for any kind of RRM problem, where a delicate trade-off between sample efficiency, robustness, and performance must be sought.
We contend that, adhering to the principle of Occam's razor, the selection of a specific solution technique for a given RRM problem hinges on the following questions: 
\begin{itemize}
\item Is it a long-term planning problem where actions now affect future rewards, or a short-term planning problem where maximizing the instantaneous reward suffices? 
\item Can the underlying model be assumed as known?
\item Can we solve the model analytically?
\item How much historical data is available?
\item Is an expert-designed policy already accessible?
\end{itemize}

With these considerations in mind, we create a decision tree for an arbitrary RRM problem that suggests the simplest technique providing an effective solution.
At one extreme there are problems where the full complexity of RL is appropriate. At the opposite extreme there are problems that can be solved directly via explicitly formulated mathematical optimization. 
In the middle, there are problems where we do not need the full generality of RL but rather \emph{some} element of learning (e.g.\ the traffic patterns and channel conditions). 
One example of a technique within this middle ground is {\em Bayesian Optimization}, where we cast the problem as single-shot optimization of a function belonging to a known family, but the exact structure of the function is not known {\em a priori}, e.g.\ because it depends on unknown aspects of the wireless environment.


There have recently been research efforts to utilize AI for the wireless physical layers, e.g., in \cite{HoydisAVV21}. We believe that the applications of AI to RRM and to the physical layers are fundamentally different, since many RRM problems are best viewed through the lens of sequential decision-making agents, which is a less appropriate paradigm for the physical layer.

\section{The model: Markov Decision Process}

Markov Decision Processes (MDP) are the standard for sequential decision planning in dynamic environments, which often arise in RRM problems. At step $t$ the environment is in \emph{state} $s_t\in \mathcal S$, encapsulating the essential information about the environment. An agent takes an action $a_t$ according to some strategy $\pi$. 
Then, the agent receives an instantaneous reward $r(s_t,a_t)$ and the state transitions to a new value with a probability dependent on the initial state $s_t$ \emph{and} action $a_t$. 
It is assumed that the state evolution is Markovian, meaning that the knowledge of past states and actions is futile in predicting the next state. 
The agent wants to maximize the long-term sum of rewards:
\begin{equation} \label{eq:obj}
    \text{objective:} \ \ \operatorname{maximize}_{\pi} \mathbb{E} \sum_{t\ge 0} \beta^t r(s_t,a_t)
\end{equation}
where $\beta\in[0,1)$ is a fixed discount factor (in practice, $\beta\approx 0.99$) and the expectation is with respect to the random reward, state evolution and strategy.
The state $s$ may not be observable; instead, the agent may perceive random \emph{observations} that are correlated with the hidden state and the action taken.



\begin{figure*}
    \centering
    \includegraphics{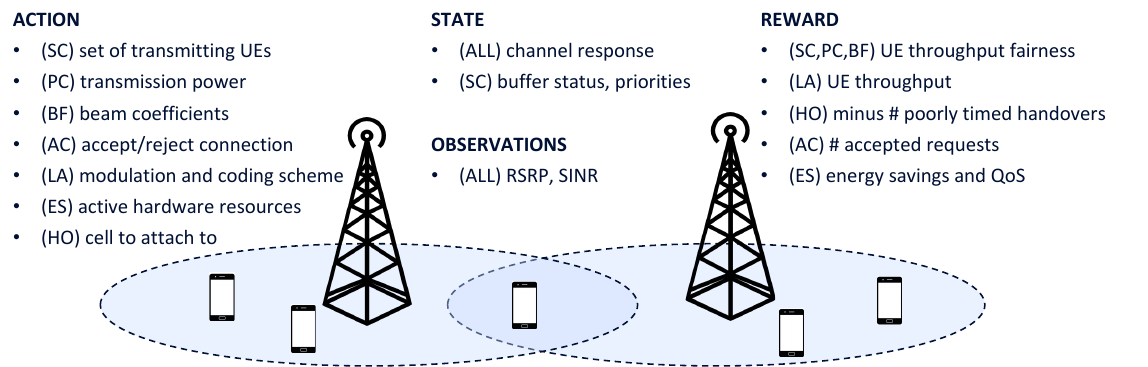}
    \caption{Pictorial illustration of the MDP formulation for various RRM use cases.}
    \label{fig:bts_and_ues}
\end{figure*}

\subsection{MDP formulation for some RRM problems} \label{sec:RRM_MDP}

In this paper we discuss how different RRM problems can be described via the MDP formalism, and yet their optimization can involve different techniques.
Next we informally characterize the MDP formulation for a range of RRM problems via the (state, action, reward) triple, along with the ``observation'' if the state is hidden, as illustrated in Figure \ref{fig:bts_and_ues}.

\subsubsection{Downlink user scheduling (SC)} At each time slot, the base station (BTS) schedules users for downlink transmission and allocates each of them to subcarriers. The goal is to achieve a fair allocation of throughput across users with respect to their channel conditions and traffic needs. \\
\noindent  \emph{State}: Channel quality, buffer status, traffic priorities, past achieved throughput for each user.\\
\noindent \emph{Action}: Selection of user(s) to be served on each subcarrier.\\
\noindent \emph{Reward}: (A fairness function of) user long-term throughput.

\subsubsection{Beamforming (BF)} The BTS steers the signal towards the intended user(s) and mitigates the interference experienced by the remaining users by controlling \emph{precoding} matrices in the digital domain and \emph{phase shifters} in the analog domain.\\
\noindent \emph{State}: Channel matrix for each user. \\
\noindent \emph{Observations}: Reference Signal Received Power (RSRP) measures for the beams previously tested or other channel statistics such as channel covariance matrices.\\
\noindent \emph{Action}: Selection of precoding matrix and analog shifters in the digital and analog domain, respectively.\\
\noindent \emph{Reward}: (A fairness function of) the throughput across users.


\subsubsection{Energy savings (ES)} The BTS uses its hardware resources parsimoniously, to minimize its energy consumption while maintaining acceptable user quality of service (QoS). \\
\noindent\emph{State}: Traffic load, user QoS, set of active hardware resources.\\
\noindent\emph{Action}:  Hardware resources (carriers, antennas) to be activated.\\
\noindent\emph{Reward}: Reduction in energy consumption at the BTS and user QoS, both to be maximized.

\subsubsection{Power control (PC)} The appropriate (uplink or downlink) transmission power allocation must be determined to ensure a good signal quality at the intended receivers, while avoiding excessive interference to non-intended ones. \\
\noindent\emph{State}: Channel matrices. \\
\noindent \emph{Observations}: SINR, across all users and BTSs.\\
\noindent\emph{Action}: Selection of power levels used for transmission.\\
\noindent\emph{Reward}: (Fairness function of) the throughput across the users.

\subsubsection{Link adaptation (LA)} Depending on the SINR, a suitable Modulation and Coding Scheme (MCS) is chosen. Overly conservative schemes unnecessarily reduce the transmission rate, while aggressive options lead to frequent retransmissions. \\
\noindent \emph{State}: SINR of a given user.\\
\noindent \emph{Action}: Selection of MCS.\\
\noindent \emph{Reward}: User throughput.

\subsubsection{Handover (HO)} Based on the RSRP measured by a user from different cells, the timing and destination of a handover (HO) to a neighboring cell is determined. Poorly timed HOs may result in degraded connectivity and radio link failures. \\
\noindent \emph{State}: RSRP at serving and neighboring cells for the user.\\
\noindent\emph{Action}: The decision of whether to perform a HO and the selection of the cell to which the user will connect next. \\
\noindent\emph{Reward}: Negation of the number of poorly timed HOs.
\subsubsection{Admission control (AC)} The network decides whether to accept or reject the incoming service request. If  resources are constrained, accepting all requests may degrade the QoS of existing services and prevent the acceptance of new ones. \\
\noindent \emph{State}: Current utilization of resources.\\
\noindent \emph{Action}: Accept, reject, or delay the incoming service request.\\
\noindent \emph{Reward}: Number of new accepted requests, QoS of existing ones, offset by a penalty on the number of rejected requests.


\begin{figure*}
    \centering
    \includegraphics[width=\linewidth]{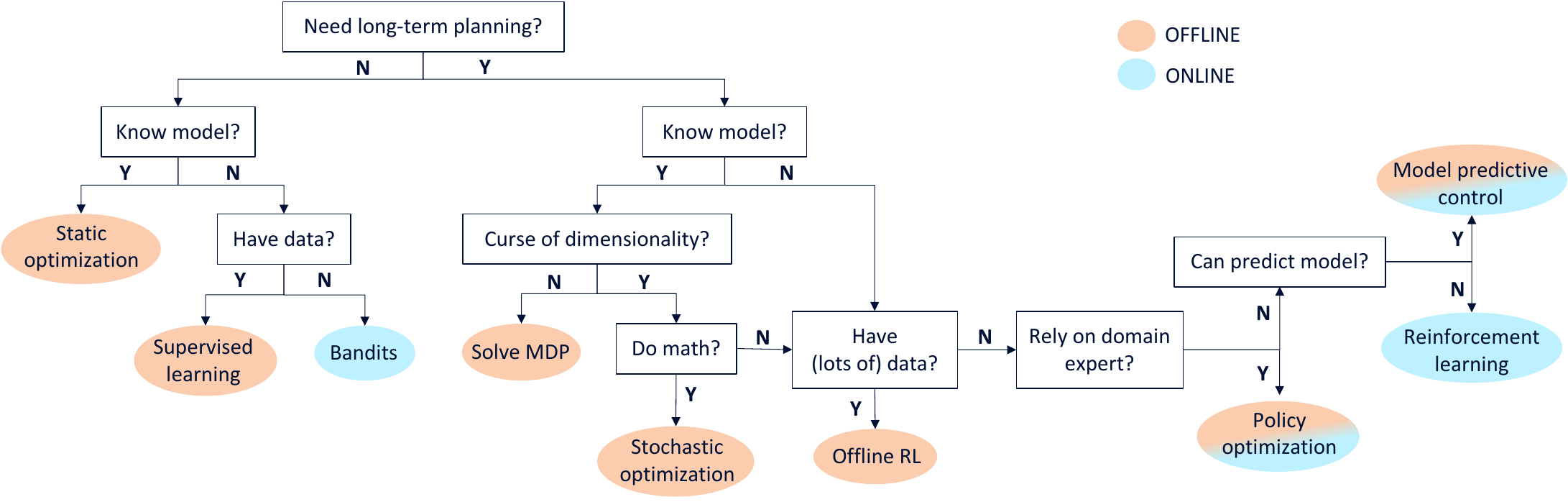}
    \caption{Proposed algorithm selection cheatsheet in sequential decision problems for RRM.}
    \label{fig:Occam}
\end{figure*}

\section{Short versus long-term planning}

A general-purpose method solving the long-term problem \eqref{eq:obj} exists under the name of \emph{Reinforcement Learning} (RL)~\cite{sutton2018reinforcement}. RL \emph{converges} to the optimal long-term solution through interaction with the environment, observation of states and rewards, and remarkably, \emph{without} requiring any prior knowledge of the underlying MDP model. However, RL is known to be sample inefficient as it relies on numerous trials and errors for learning \cite{dulac2021challenges}. This is unacceptable in many practical RRM scenarios, especially those operating on the network layer and above, making decisions on slow time scales, in the order of tens of seconds or minutes. 

To address this concern it is useful to follow the general Occam's razor guidelines: when faced with competing options, the \emph{simplest} one should be preferred. 
In our scenario, this translates to identifying the \emph{minimal} set of features that make the sequential decision problem \emph{hard}, and solve for them.
Therefore, in this paper we provide a set of rules, depicted in Figure \ref{fig:Occam}, that can be followed to customize the solution method to the RRM problem at hand.

We now discuss the initial coarse distinction one should make, between short and long-term planning algorithms.  

\subsection{Endogenous state evolution $\Rightarrow$ Long-term planning.} 
When taking decisions sequentially, it is typically important to assess the long-term consequences of the current action. This principle is evident in the MDP formalism, where \emph{long-term planning} arises from the fact that selecting an action influences the probability distribution of the subsequent state. Hence, the agent may want to choose an action that strikes a trade-off between instantaneous reward and future trajectory of states, as they are correlated with future rewards.

\subsection{Exogenous state evolution $\Rightarrow$ Short-term planning}
Assume now that the states evolve exogenously, i.e., the agent cannot control the state trajectory, then the agent does not need to consider the future when taking an action. Thus, the agent's task simplifies to greedily maximizing the instantaneous reward alone. 
We remark that the converse does not always hold: the myopic strategy can be optimal even if the state evolution is endogenous, under some conditions \cite{sobel1981myopic}.

\subsection{Short versus long-term planning in RRM use cases} 

Based on the short- and long-term planning differentiation above, only some RRM use cases (may) require long-term planning.

\subsubsection{SC, AC} In scheduling (SC) and admission control (AC), the state evolution is affected by the agent's action. In SC, the user buffer status and achieved throughput depend on the subset of users scheduled for transmission. In AC, deciding to accept more requests can prevent the acceptance of future requests. 
Accepting all incoming requests without considering resource limitations may lead to congestion, requiring preemptive delay or rejection of new requests.

\subsubsection{ES, HO} A greedy policy may seem optimal for energy savings (ES) and handover (HO) optimization use cases, activating only essential hardware components for ES and connecting to the cell with the highest RSRP for HO. 
However, in both use cases, the action may materialize with some inertia, prompting the need for long-term planning. In HO, users may delay switching to a neighboring cell until it consistently outperforms the current one in RSRP to avoid frequent reconnections. In ES, hardware components should only be shut down when predicted QoS remains acceptable with high confidence, considering the delays associated with component reactivation.

\subsubsection{PC, BF, LA} According to the MDP formulations above, short-term planning is sufficient for the use cases on Power control (PC), Beamforming (BF) and Link adaptation (LA). 
The action has no impact on the evolution of the state, consisting of the user channel matrix. 
Yet, especially in the PC case, it is customary to introduce some inertia in the choice of actions to mitigate erratic control behavior, which prompts the use of long-term planning, as discussed later on.



\begin{figure*}
    \centering
    \includegraphics[width=\linewidth]{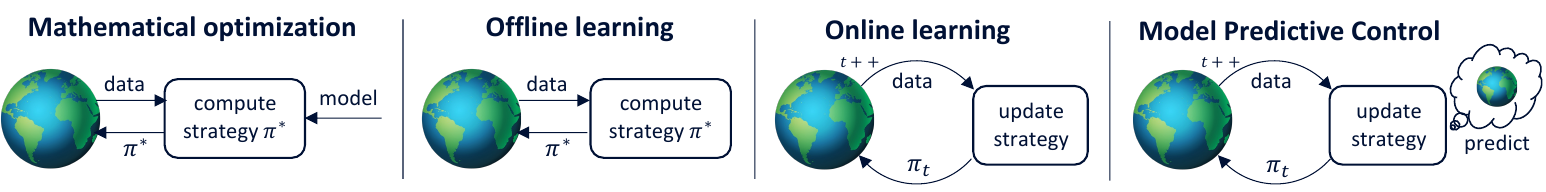}
    \caption{Comparison of various techniques regarding their interaction with the live system.}
    \label{fig:4figs}
\end{figure*}

\section{Short-term optimization}

We start navigating through the algorithm diagram for sequential problems in RRM shown in Figure \ref{fig:Occam}. 
In this section we assume that the state evolves exogenously with respect to the agent's strategy, hence short-term planning is sufficient. 

\subsection{Model known: Static optimization} \label{sec:model_known}

We can sometimes assume that the reward function, mapping state and actions to rewards, is available in advance to the agent. 
In conjunction with the short-term assumption, one can maximize the long-term reward \eqref{eq:obj} by directly maximizing the instantaneous expected reward using classic \emph{static} optimization techniques, ranging from combinatorial to continuous, and possibly convex, methods. 
This ``lucky'' case emerges in BF and PC, under the assumption that the user channel matrix, i.e., our MDP state, is known by the agent, typically the BTS. 
Here, the state evolves independently of the agent's actions, solely based on environmental scattering characteristics. 

In \textbf{BF} for Multiple Input Multiple Output (MIMO) antenna systems, precoding vectors at BTS (which represent the action), can be chosen to maximize a reward equal to the downlink sum rate across users, subject to a constraint on the transmitted power. 
If the reward is expressed via the Shannon capacity formula as a function of the SINR of each user, then the optimal action is computed analytically as the Minimum Mean Square Error (MMSE) precoding matrix. 
In \textbf{PC}, a challenge is allocating power across independent point-to-point channels to maximize the sum of achievable rates while adhering to a constraint on total transmitted power. 
This results in \emph{water-filling} solutions, derived from the Karush-Kuhn-Tucker conditions of the constrained convex problem. 


\subsection{Model unknown: Learning}  \label{sec:bandits}

When the reward function is not available in closed form to the agent or the state cannot be fully observed, then mathematical optimization in the form discussed in Section \ref{sec:model_known} is no longer an option. 
Then, two main options emerge.

\subsubsection{Supervised learning} 
Assume that the agent has collected a historical dataset reporting (state $s_t$, action $a_t$, reward $r(s_t,a_t)$) triples. Then, the agent may apply supervised learning and train a function approximator, e.g., a deep neural network, to infer the reward function or the optimal action, directly.
The optimal action in each state can be found by maximizing the inferred reward function via gradient descent.
This approach can be used in analog \textbf{BF}, where a recurrent neural network is trained, e.g., to predict the angle of departure for a given user \cite{elbir2023twenty} given previous RSRP measurements.


\subsubsection{Bandits} 
A different option is to leverage historical data solely to establish prior information about the reward function, and refine the model by interacting with the live system.

Optimizing a partially unknown reward function based on an endogenous state can be achieved with \emph{contextual multi-armed bandits} (CMAB). There exist different flavors of CMAB, ranging from the case where the reward is uncorrelated across states and actions, akin to the classic multi-armed bandit (MAB) problem, or a smooth function of both states and actions, typically addressed via Bayesian optimization (BO). 
In BO, the probabilistic belief about the unknown reward is updated after new observations are gathered, and the next point is chosen to solve the exploration \emph{vs.} exploitation dilemma. 
Following an empirical Bayes approach, historical data can be exploited to initialize the Bayesian prior.
The smoothness of the reward facilitates learning the function shape near observed (state, action) pairs, increasing algorithm convergence.
In RRM scenarios, it is typical to assume the reward as a smooth function of the state, particularly when the state comprises continuous measures such as traffic load, user SINR, or simply time. 
Actions, however, can be continuous or discrete. The former scenario is typical in the \textbf{PC} use case. 
For instance, in \cite{maggi2021bayesian} the uplink PC parameters are optimized via BO, assuming that the reward (a fairness function of the throughout across the users) is smooth with respect to the action. 
A similar approach can be applied to analog \textbf{BF}. In contrast to digital BF, the channel matrix may be \emph{hidden} in analog BF, due to overhead concerns arising from the large number of antennas. Then, the reward function is also unknown to the agent. Yet, the agent can \emph{observe} the associated RSRP and assume that it varies smoothly with the beam's main lobe direction and time, and optimize it accordingly in an online fashion, as in \cite{maggi2023tracking}.

In \textbf{ES}, actions can represent, e.g., the kind of hardware resource to be used for a certain task. The agent can use CMAB to decide whether to allocate the new computing task to energy-efficient CPUs or to high-performing hardware accelerators, aiming to minimize energy consumption without breaching processing delay constraints, as in \cite{ayala2023risk}. 
In \textbf{LA}, the MCS (our action) guaranteeing a certain block error rate (BLER) for the current SINR (state) is sought. The optimal MCS could be computed via simulation for each value of SINR, creating a lookup table. 
Yet, two issues arise: real systems differ from simulations and the user's SINR reports are not reliable. Thus, an inner/outer-loop method called ILLA/OLLA can be used, where ILLA computes the optimal MCS from a look-up table and OLLA corrects it by observing the user's ACK/NACK. 
A more direct approach estimates explicitly the BLER probability for each MCS. Then, the agent chooses the MCS maximizing the achieved throughput (i.e., the reward), which is the (known) data rate multiplied by the (estimated) 1-BLER via Thompson sampling, as in \cite{saxena2021reinforcement}.

\section{Long-term optimization}

We continue to navigate our diagram for RRM algorithm selection in Figure \ref{fig:Occam} and deal with problems requiring long-term planning, i.e., for which the myopic strategy maximizing the instantaneous reward is sub-optimal in the long-run.

\subsection{The lucky case: MDP known and tractable}

If the MDP underlying our RRM use case is \emph{known} and \emph{tractable}, then the \emph{optimal} strategy can be computed using traditional techniques such as policy or value iteration and linear programming. 
Yet, in practical scenarios, the \emph{curse of dimensionality} often thwarts us from employing such solution methods. This occurs when states and/or actions are continuous and, even worse, multi-dimensional, making naive discretization an impractical approach.

\subsection{Ad-hoc stochastic optimization techniques}

In some cases it is still possible to compute the optimal solution for an MDP by using \emph{ad-hoc} stochastic optimization techniques.
In \textbf{SC}, to achieve throughput proportional fairness (PF), the agent must serve the user with the highest ratio of current spectral efficiency to past achieved throughput \cite{stolyar2005maximizing}. 

A similar technique known as Drift-Plus-Penalty (DPP) has been developed for network utility maximization and applied, among others, to \textbf{ES} use cases \cite{neely2006energy}, where the network energy expenditure is minimized while ensuring the stability of all user queues through adjustments in the transmission power.

Remarkably, neither PF nor DPP require knowledge of the state transition rule; they only assume knowledge of the reward function. Moreover, they do \emph{not} explicitly solve a multi-step problem; instead, they maximize a suitably \emph{modified} reward function that depends only on past measures.

\textbf{AC} is another fertile playground for stochastic optimization. 
For instance, one can prove that when requests have different priorities, the optimal admission strategy is typically in the form of a \emph{trunk reservation} one, where requests are accepted only if the available bandwidth exceeds a threshold that decreases with the priority of the request \cite{altman2002applications}.


\begin{figure}
    \centering
    \includegraphics[width=\linewidth]{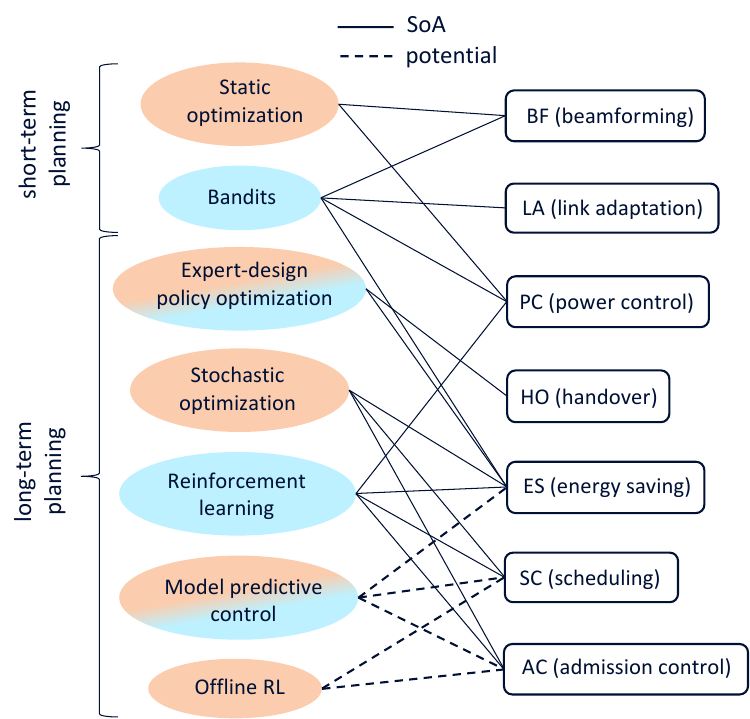}
    \caption{Mapping between optimization techniques for sequential planning and RRM use cases.}
    \label{fig:map}
\end{figure}

\subsection{Exploit historical data: Offline Reinforcement Learning}

We briefly discuss a burgeoning field, known as \emph{Offline Reinforcement Learning} \cite{levine2020offline} that becomes relevant if the agent has collected a large amount of data reporting information on state, action and collected reward.
Similar to supervised learning (SL) in short-term planning, in offline RL the agent cannot interact with the live system and must rely solely on offline data to optimize its policy. 
At its core, offline RL faces the challenge of making counterfactual arguments  (i.e., considering what would have happened if a different action had been chosen) without the ability to actually make different decisions. This challenge is compounded by the fact that a change in action can impact the state evolution over multiple time steps. 
This issue is less prevalent in SL, where actions only influence current rewards and the effect of unobserved actions can be inferred via interpolation.  
Despite the promising potential of offline RL, significant advancements are still needed to ensure its safe application in real-world systems.

\subsection{Rely on domain-expert designed strategies} \label{sec:domain_expert} 

It is common for RRM domain experts to craft \emph{classes} of parameterized policies. 
A prominent example is Mobility Robustness Optimization (MRO), within \textbf{HO}. Here, connecting greedily to the cell offering the highest instantaneous RSRP may result in unstable behavior for two main reasons: noisy RSRP measurements and the non-instantaneous nature of the HO procedure. Thus, an incorrect HO decision cannot be immediately reversed if detrimental. 
To address this, 3GPP has standardized policies, that we interpret as MDP parameterized policies, where the state indicates the number $n$ of consecutive measurements where the RSRP of neighboring cells has exceeded the RSRP of the primary cell, adjusted by a hysteresis value. The HO action is made only when $n$ surpasses a \emph{time-to-trigger} threshold. 
If such parameters are properly configured, then the MRO policy can prevent radio link failures and minimize ping-pongs. 

Expert policies also emerge in \textbf{ES}, where a reasonable policy involves activating the fewest hardware components necessary 
to guarantee that the average resource utilization in a sector remains within a lower and an upper threshold. These thresholds act as policy parameters, as in \cite{maggi2023energy}. 

To optimize the parameters of such policies, the agent has two main options.
The former is agnostic to the underlying MDP structure, and optimizes the parameters in a black-box manner: the agent converges to the optimal value by experimenting with different values and estimating the associated long-term reward. 
Derivative-free methods are suited for this case, e.g., Nelder-Mead and BO, once again.
These techniques solve a long-term planning \emph{indirectly}, but really they are designed for static optimization.
The latter option involves using policy gradient techniques borrowed from the RL literature. These compute the gradient of the long-term reward with respect to the parameters by leveraging the underlying MDP model and performing gradient ascent. 

\subsection{Model predictive control} \label{sec:mpc} 

Model predictive control (MPC) is often overlooked in the field of telecommunications, despite being extensively studied in optimal control and widely applied in various industries. 
MPC proceeds as follows. First, the future trajectory of states, based on the applied actions, is \emph{predicted} at the current step. In RRM, this would entail predicting, e.g., future channel quality, incoming requests and mobility patterns.
Then, the optimal strategy is computed over a finite time horizon by making certain assumptions on the state dynamics and reward structure \cite{grne2013nonlinear}. Finally, only the optimal action in the \emph{current} state is applied. This process is repeated at each step.

MPC yields good results in practice even if the state predictions are poor and \emph{deterministic}. The reason is intuitive: predictions are frequently updated alongside the associated policy, letting MPC \emph{adapt} to changes in the environment. 

The benefits of MPC also uncover its weaknesses: computational complexity may be significant, arising from the state prediction and, primarily, the computation of the optimal policy at each step. Figure \ref{fig:4figs} stylizes the difference between MPC and more classic offline and online learning methods.

\subsection{Reinforcement Learning}

When the underlying MDP model is unknown, a domain expert policy is not trusted or the state evolution cannot be reliably predicted, then Reinforcement Learning (RL) emerges as the necessary option. 
Several applications of RL to RRM problems are documented in the literature. 
RL is effective for \textbf{SC} when the analytical PF metric falls short, e.g., when constraints are imposed on the transmission delay for each user, as in \cite{hu2022effective}. 
In \textbf{AC}, RL becomes a viable option when the model cannot be treated analytically and the structure of the optimal admission policy remains unknown. This situation arises, e.g., when considering requests with diverse requirements concerning delay, CPU usage and bandwidth. 

In the context of uplink \textbf{PC}, it is customary to introduce action inertia by limiting the variation of uplink power to small incremental amounts, thus preventing erratic behaviors. In such cases, long-term planning becomes necessary and RL can accommodate this requirement, as shown in \cite{neto2021uplink}.
Finally, in \textbf{ES}, RL can tackle complex scenarios where the action is highly dimensional and associated to intricate trade-offs. This is evident when the action entails the shutdown of multiple frequency layers and BTSs, each characterized by different QoS, energy efficiency and reactivation latency. 
Crafting a reasonable policy for such scenarios is challenging for a domain expert, making RL an appropriate choice to learn an effective policy from scratch.

\section{A more elaborate example: Beamforming}

\begin{figure}
    \centering
    \includegraphics[width=\linewidth]{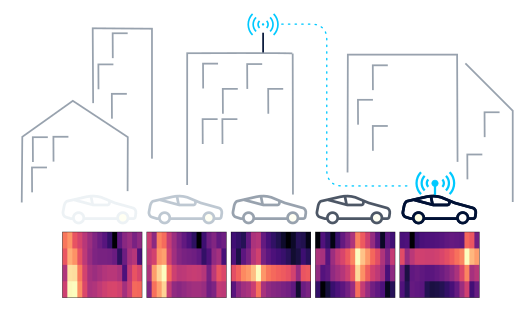}
    \caption{User RSRP across beams and over time}
    \label{fig:bf1}
\end{figure}

We conclude with a more detailed example on analog \textbf{BF} in millimeter wave. 
The beam with the maximum \emph{instantaneous} RSRP is sought, for each user and at every time step. Therefore, long-term planning is unnecessary. The channel matrix, i.e., the state, is hidden, hence ``static optimization'' is not possible. After beam selection, the user feeds the RSRP, i.e., the observation, back to the BTS. Figure \ref{fig:bf1} shows the time evolution of RSRP across beams. 

\begin{figure}
    \centering
    \includegraphics[width=\linewidth]{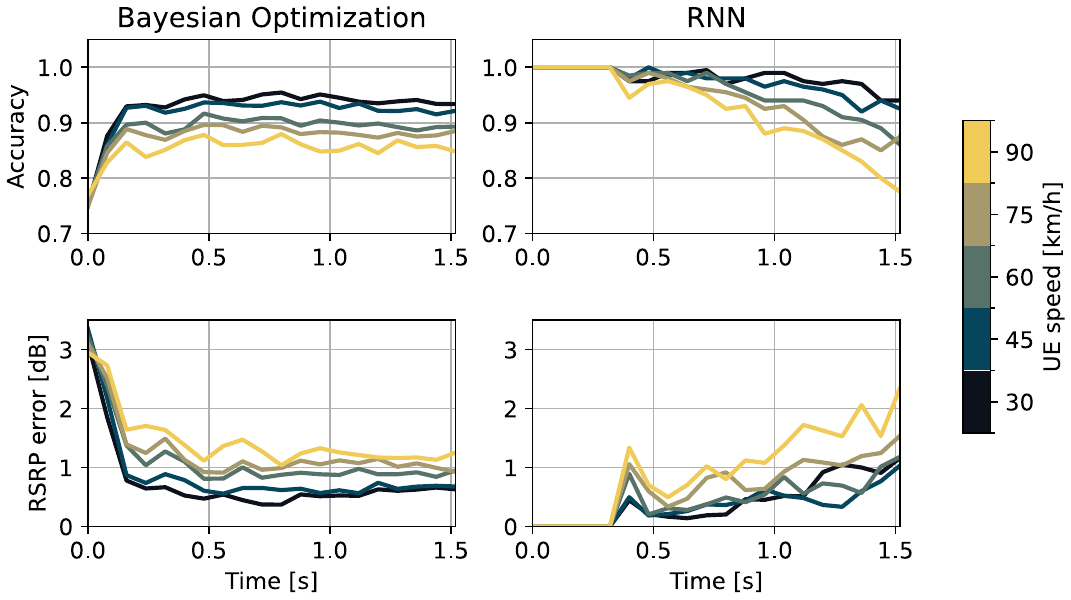}
    \caption{Accuracy (probability of finding the optimal beam) and error (selected beam versus the optimal beam) showing different performance characteristics of BO (left column) and RNN (right column) for varying UE speeds.}
    \label{fig:bf2}
\end{figure}

Measuring RSRP on too many beams results in excessive overhead, while too few measurements can be detrimental for beam calibration and, eventually, throughput.
According to the diagram in Figure \ref{fig:Occam}, two options are available here: ``supervised learning'' and ``bandits''.
Within the former category, a Recurrent Neural Network (RNN) can be trained on historical data to predict a sequence of best beams for the user \cite{dehkordi2021adaptive} given previous RSRP measurements, exploiting their correlation over time. 
The trained RNN is then deployed at the BTS. 
Updating the RNN model at the BTS during live operations is arguably a delicate operation, although it may be required if traffic patterns change.
Additionally, performance degrades over the sequence of predicted beams, requiring expensive periodic gathering of RSRP measurements (input sequences).
To address this concern we can turn to the ``bandit'' category, where the model is continually updated at the BTS. Bayesian optimization (BO) helps choose the subset of beams to measure at each step, maximizing expected RSRP while penalizing the number of beams used to account for overhead. Leveraging spatial and temporal correlations, we update beliefs for RSRP at other unmeasured beams and times using a smooth statistical model.  Fig.~\ref{fig:bf2} illustrates this process. 
However, performing these operations for each user independently can impose a significant computational burden at the BTS. A more detailed analysis of these methods' pros and cons is explored in \cite{maggi2023tracking}.





\section{Conclusions}

The algorithm cheat-sheet for RRM proposed in this paper is by no means exhaustive. 
For instance, one may wonder what happens when, in the decision tree of Figure \ref{fig:Occam}, we are allowed to provide non-binary answers.
An intriguing scenario arises when we can derive the theoretically optimal solution for a given model describing the problem at hand but cannot fully rely on the model itself.  
One approach is to adjust the theoretical solution by introducing additive and/or multiplicative parameters. 
These parameters can be fine-tuned on the live system either in black-box manner, as discussed in Section \ref{sec:domain_expert}, or via MPC, as in Section \ref{sec:mpc}. 

In this paper we advocate for a broad range of optimization techniques to be considered for RRM use cases. We argue that the selection of the algorithm should follow the guidelines outlined in Figure \ref{fig:map}. 
We encourage new exciting applications of MPC and offline RL to RRM use cases with a strong long-term planning component, such as SC and AC. Special efforts should be dedicated to devising algorithms that respect the tight latency constraints.
For the scenarios where short-term planning suffices, such as LA and BF, we suggest further investigation of learning techniques such as (contextual) bandits and their variants, discussed in Section \ref{sec:bandits}.

\section*{Acknowledgments} 
The authors express their gratitude to their colleagues Claudiu Mihailescu, Suresh Kalyanasundaram, Alvaro Valcarce Rial and Jun He for inspiring discussions over the years.


%
%
%
%

\end{document}